\documentclass[aps,prd,twocolumn,showpacs,superscriptaddress,groupedaddress]{revtex4}  
\usepackage{graphicx}  
\usepackage{dcolumn}   
\usepackage{bm}        
\usepackage{amssymb}   
\usepackage{threeparttable} 
\usepackage{subfig}  
\usepackage{changepage}

\begin{document}

\title{Cosmology from cross correlation of CMB lensing and galaxy surveys}
\date{\today}

\author{R. Pearson}
\address{Department of Physics \& Astronomy, University of Sussex, Brighton BN1 9QH, UK}
\address{Kavli Institute for Particle Astrophysics and Cosmology, SLAC, 2575 Sand Hill Road, Menlo Park, CA 94025, USA}

\author{O. Zahn}
\address{Department of Physics, University of California, Berkeley, USA}
\address{Lawrence Berkeley National Laboratory, Berkeley, CA, USA}

\begin{abstract}
In recent years cross correlation of lensing of the Cosmic Microwave Background (CMB) with other large scale structure (LSS) tracers has been used as a method to detect CMB lensing.  Current experiments are also becoming sensitive enough to measure CMB lensing without the help of auxiliary tracers.  As data quality improves rapidly, it has been suggested that the CMB lensing-LSS cross correlation may provide new insights into parameters describing cosmological structure growth.  In this work we perform forecasts that combine the lensing potential auto power spectrum from various future CMB experiments, the galaxy power spectrum from galaxy surveys, as well as the cross power spectrum between the two, marginalizing over a number galactic and non-galactic cosmological parameters. 
We find that the CMB lensing-LSS cross correlation contains significant information on parameters such as the redshift distribution and bias of LSS tracers.
We also find that the cross correlation information will lead to independent probes of cosmological parameters such as neutrino mass and the reionization optical depth. 
\end{abstract}

\maketitle

\section{Introduction}
In the past few years CMB lensing science has arisen as a promising new probe of cosmology.  CMB photons are deflected by matter fluctuations between the surface of last scattering and today. The effect is sensitive to the geometry of the Universe (e.g., dark energy and its evolution) and the shape of the matter power spectrum (affected by e.g., massive neutrinos).  Since CMB lensing probes all matter, it should be correlated with galaxy surveys observing the large scale structure of the Universe. The cross correlation between  CMB lensing and such large scale structure surveys should aid in simultaneously measuring galactic and non-galactic cosmological information.  There is already a wealth of data available for such analysis, and upcoming surveys will offer even better opportunities for cross correlation science.  A few examples of work to date include: CMB lensing $\times$ radio galaxies using WMAP and NVSS \cite{Smith:2007rg,Feng:2012uf}, CMB lensing $\times$ LRGs $\times$ quasars $\times$ radio galaxies using WMAP, NVSS and SDSS \cite{Hirata:2008cb}, CMB lensing $\times$ optical $\times$ IR using SPT, BCS, WISE and Spitzer \cite{Bleem:2012gm}, CMB lensing $\times$ quasars using ACT, and SDSS \cite{Sherwin:2012mr}, CMB lensing $\times$ sub-millimetre wavelength galaxies using SPT and Herschel/SPIRE \cite{Holder:2013hqu}.

Both CMB temperature and polarization are being measured by modern experiments.  At the detector level, the polarization measurement is in terms of the Stoke's parameters Q and U, but it is usually useful to consider CMB polarization in terms of even and odd parity modes, the E- and B-mode respectively.  The temperature anisotropy has been measured well down to arc minute scales (e.g. \cite{Jones:2005yb,Hinshaw:2012aka,Planck:2013kta,Calabrese:2013jyk,Story:2012wx}), E-mode polarization has been detected by several groups (e.g. \cite{Leitch:2004gd,Montroy:2005yx}), and the conversion of E-mode to B-mode polarization by lensing has been observed \cite{Hanson:2013hsb}.  An observation of large scale B-modes extra to lensing and foreground components could be direct evidence of a primordial gravitational background, and used to discern inflationary theories. Even without the B-mode though, observations of E-mode polarization allow improved constraints on cosmological parameters compared to temperature information alone, since they are sensitive to the epoch of reionization as well as recombination physics.  To obtain the CMB lensing signal, the quadratic estimator method of \cite{Hu:2001kj} is usually used.  The CMB lensing potential can be reconstructed from various combinations of T-, E- and B-mode measurements.  So far the lensing power spectrum has been measured using the temperature anisotropy observed by the Atacama Cosmology Telescope (ACT) \cite{Das:2011ak}, \cite{Das:2013zf}, the South Pole Telescope (SPT) \cite{vanEngelen:2012va}, and most recently the Planck satellite \cite{Ade:2013tyw}.  
Lensing also affects the peaks and troughs in the angular power spectra (e.g., \cite{Keisler:2011aw,Story:2012wx}).
Finally, CMB lensing  mixes primordial E-modes into lensing-specific B-modes. This allows reconstruction of the CMB  lensing potential with much smaller variance than for the temperature modes, because no primordial B-modes are expected on small scales.

A Fisher matrix calculation marginalizing over various cosmological parameters can be used to estimate the constraining power of current and future data sets.  In this work, different CMB and galaxy survey sensitivities are explored with the goal of showing how well cosmology can be constrained in the presence of uncertainties in the galaxy physics, and vice-versa.  In particular, we are interested in how well the parameters are constrained using the CMB lensing - galaxy cross correlation. 
Seven non-galactic parameters are included in the analysis with the following fiducial values: physical baryon fraction ($\Omega_{\rm b}h^{2}$=0.02258), physical cold dark matter fraction ($\Omega_{\rm CDM}h^{2}$=0.1109), dark energy fraction ($\Omega_{\lambda}$=0.734), physical neutrino mass fraction ($\Omega_{\nu}h^{2}$=0.002), optical depth ($\tau$=0.088), spectral index ($n_{\rm s}$=0.963), scalar amplitude ($A_{\rm s}$=2.43 x10$^{-9}$); and up to three galactic parameters: the linear galaxy bias (b), the mean of the galaxy redshift distribution ($\mu$), and the standard deviation of the galaxy redshift distribution ($\sigma$).

Previous work has already shown that cross-correlation of structure and CMB lensing can constrain various biases, and thus provide useful information for cosmology\cite{Vallinotto:2011ge,Vallinotto:2013eva,Rhodes:2013fyq}.   In particular, CMB lensing can break degeneracy between the matter power spectrum normalisation and systematic multiplicative biases in shear measurements \cite{Vallinotto:2011ge}, and can constrain the linear galaxy bias \cite{Vallinotto:2013eva}.  Our work complements the latter analysis by allowing redshift distribution parameters to also vary.

Section~\ref{method} introduces the Fisher methodology used and outlines the derivation of our theoretical power spectra.  In section~\ref{data} we describe some current and future CMB and galaxy surveys as well as our data cuts.  Results on cosmological parameters are given in section~\ref{results} and we conclude in section~\ref{conclusion}.

\section{Method}
\label{method}
For a given cosmological model described by parameters $\lambda_i$ and future experiment with known specifications, one can approximate the likelihood in the vicinity of the best-fit (fiducial) model as a multivariate Gaussian with the Hessian given by the Fisher matrix in terms of the theoretical power spectra $C_l$ as 
\begin{equation}
\label{Fish}
F_{\alpha\beta}=\sum_{l}\frac{1}{(\delta C_{l})^2}\frac{\partial C_{l}}{\partial\lambda_{\alpha}}\frac{\partial C_{l}}{\partial\lambda_{\beta}}
\end{equation}
The uncertainty $\delta C_{l}$ is given by:
\begin{equation}
\label{error}
\delta C_{l}=\sqrt{\frac{2}{(2l+1)f_{\rm sky}}}[C_{l}+C_{l}^{\rm noise}]
\end{equation}
Where the first term in the bracket accounts for cosmic variance in the signal and $C_{l}^{\rm noise}$ is the associated experimental noise.  The factor $f_{\rm sky}$ accounts for the fraction of sky covered by the experiment.  The error on a given parameter $\lambda_{\alpha}$ is $\sqrt{(F^{-1})_{\alpha\alpha}}$.  If there is some prior constraint on the error of a parameter $\sigma_{\rm prior}$, this information is added in the Fisher matrix via $F_{\alpha\alpha}=F_{\alpha\alpha}+\frac{1}{\sigma^{2}_{\rm prior}}$.

In this analysis we form two Fisher matrices.  The first includes auto and cross correlations from CMB temperature and polarization: TT, EE, TE.  The second includes auto and cross correlations of the CMB lensing potential and 2-dimensional galaxy power spectrum: $\phi\phi$, GG, $\phi$G. When considering multiple power spectra (non-diagonal covariance), the Fisher matrix is formulated as shown in~\cite{Perotto:2006rj} via:
\begin{equation}
\label{ }
F_{\alpha\beta}=\sum^{l_{\rm max}}_{\rm l=2} \sum_{\rm PP',QQ'} \frac{\partial C_{l}^{\rm PP'}}{\partial \lambda_{\alpha}}(\mathbb{C}_{l}^{-1})_{\rm PP'QQ'}\frac{\partial C_{l}^{\rm QQ'}}{\partial \lambda_{\beta}}
\end{equation}
In the first Fisher matrix $PP',QQ' \in {TT, EE, TE}$, whilst in the second Fisher matrix $PP',QQ' \in {\phi\phi, GG, \phi G}$. $\mathbb{C}_{l}$ is the 3$\times$3 covariance matrix between the different power spectra, and the elements are made up of the appropriate $(\delta C_{l})^{2}$.  In the CMB temperature and polarization case, the $C_{l}^{\rm noise}$ is taken to be the standard Gaussian random detector noise as in~\cite{PhysRevD.52.4307}.  For the CMB lensing, $C_{l}^{\rm noise}$ is the N0 bias as in~\cite{Hu:2001kj}, and for the galaxy power spectrum it is simple shot noise.  The two Fisher matrices are added together for the final constraint, and we neglect small correlations between CMB temperature and lensing (T$\phi$) caused by the integrated Sachs-Wolfe (ISW) effect (e.g. \cite{Ade:2013dsi}).  

The theoretical auto and cross CMB lensing and galaxy power spectra are approximately:
\begin{equation}
\label{ }
C_{L}^{\phi\phi}\sim\int d\chi (K^{\phi}_{L}(\chi))^{2}P(k=L/\chi,\chi)
\end{equation}
\begin{equation}
\label{ }
C_{L}^{\rm gg}\sim\int d\chi (K^{\rm g}_{L}(\chi))^{2}P(k=L/\chi,\chi)
\end{equation}
\begin{equation}
\label{ }
C_{L}^{\phi g}\sim\int d\chi K^{\phi}_{L}(\chi)K^{\rm g}_{L}(\chi)P(k=L/\chi,\chi)
\end{equation}
Where $\chi$ is conformal distance and P(k,$\chi$) is the 3D matter power spectrum at wavenumber k and conformal look back time $\chi$.  The kernels $K^{\phi}_{L}(\chi)$ and $K^{\rm g}_{L}(\chi)$ fold in information about the lensing and the galaxy dynamics respectively:
\begin{equation}
\label{eq:Kphi}
K^{\phi}_{L}(\chi)=-\frac{3\Omega_{m}H_{0}^{2}}{L^{2}}\frac{\chi}{a}(\frac{\chi_{*}-\chi}{\chi_{*}\chi})
\end{equation}
\begin{equation}
\label{eq:Kg}
K^{\rm g}_{L}(\chi)=\frac{dN}{dz}\frac{dz}{d\chi}\frac{b(z)}{\chi}
\end{equation}

Equation~\ref{eq:Kphi} depends on non-galactic cosmological parameters, whereas Equation~\ref{eq:Kg} depends on both galactic and non-galactic cosmological parameters, since it includes the bias and redshift distribution.

\section{Data}
\label{data}
The derivatives used in the Fisher matrix are calculated using the theoretical power spectra.  CMB temperature, CMB polarization, CMB lensing and galaxy power spectra were calculated for the given parameter choice using CAMB sources \cite{Challinor:2011bk}.  To obtain the galaxy auto correlation (GG) and galaxy-CMB lensing cross correlation ($\phi$G), we used a slightly modified version of the code with source and lensing type windows.

Table~\ref{CMBspecs} shows the CMB experimental scenarios we consider in this work.  The current type experiment is based on the {\it Planck} satellite, for which the blue book values agree reasonably well with the satellite's performance so far. The 3rd generation experiment is a survey with higher angular resolution (1 arcmin FWHM) covering 10\% of the sky. We verified that a three times larger beam would not qualitatively change any of the results.   The 4th generation experiment has the same angular resolution as the 3rd generation experiment but covers 50\% of the sky.  The noise in the CMB temperature and polarization is shown in the table, and the CMB lensing noise is derived from these values.  The 3rd generation is representative of  surveys already taking data or soon to begin observations, such as SPTpol \cite{Austermann:2012ga}, ACTpol \cite{Niemack:2010wz} and POLARBEAR \cite{Kermish:2012eh}.  The 4th generation is representative of larger sky area surveys planned for the near future, such as the POLARBEAR extension, the Simons-array. 

\begin{figure}
\begin{center}
\hspace*{-0.2cm}
\includegraphics[width=8.8cm]{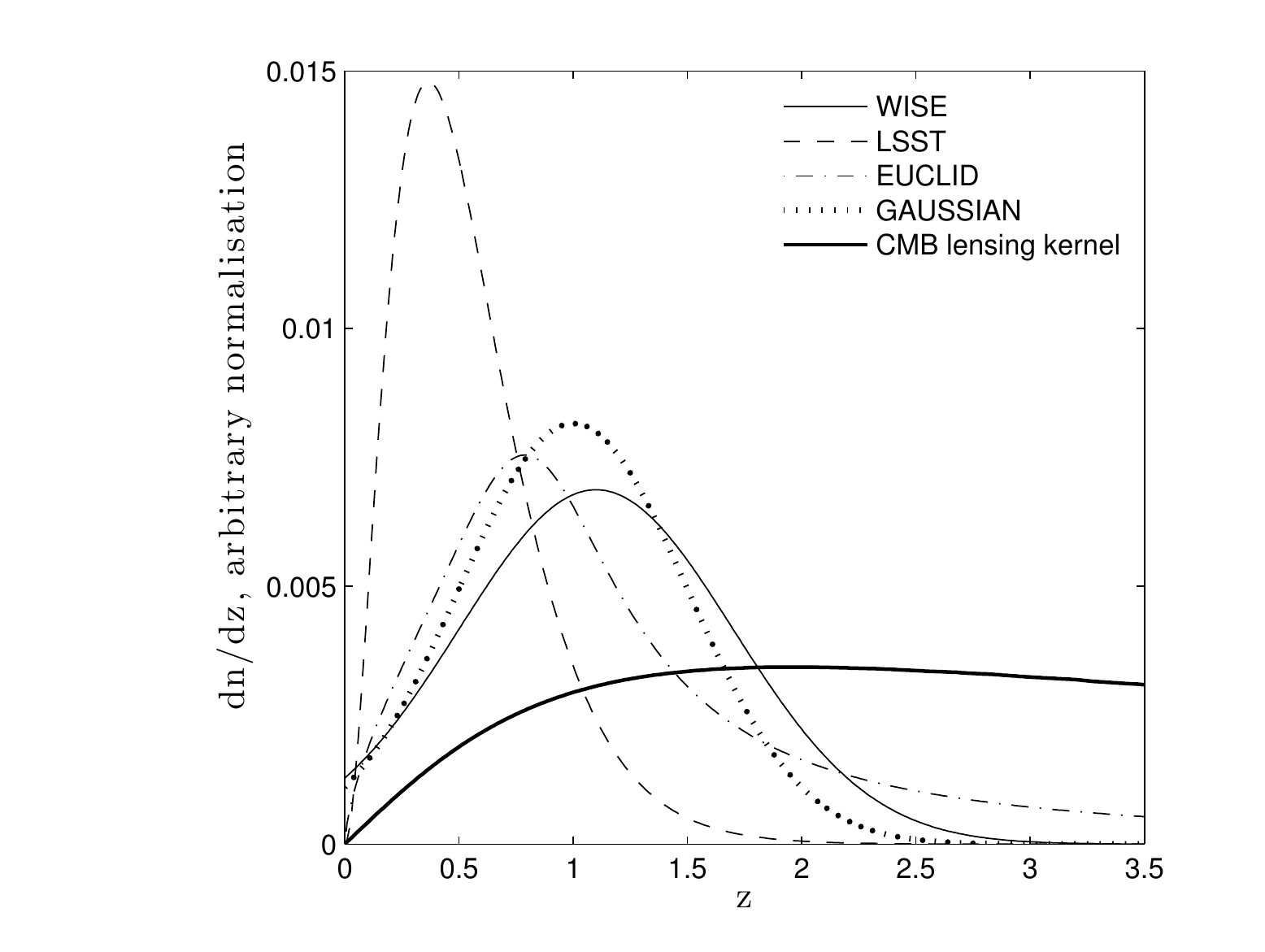}
\caption{Redshift distributions for current and up coming galaxy surveys compared to the CMB lensing redshift kernel.  Redshift distributions shown for the WISE, LSST and EUCLID surveys are taken from~\cite{Geach13}, ~\cite{Abell:2009aa} and~ \cite{Boldrin:2012ke} respectively.  The Gaussian model has $\mu=1$, $\sigma=0.5$.  }
\label{fig:dndz}
\end{center}
\end{figure}

Table~\ref{GALspecs} shows experimental scenarios for 3 galaxy surveys.  The Wide-field Infrared Survey Explorer (WISE) is a satellite experiment which observed the whole sky in the mid-infrared.  The four observing bands are centered at 3.4, 4.6, 12 and 22 $\mu$m \cite{Wright:2010qw}.  WISE has already been used in cross correlation with various data sets such as WMAP \cite{Goto:2012yc}, {\it Planck} \cite{Ade:2013tyw} and SPT \cite{Geach13}.  The Large Synoptic Survey Telescope (LSST) is a planned 8.4 meter optical ground based telescope which will observe from Cerro Pach\'{o}n, Chile.  The predicted redshift distribution from the LSST science book \cite{Abell:2009aa} is:
\begin{equation}
\label{LSST}
P(z)=\frac{1}{2z_{0}}(\frac{z}{z_{0}})^{2}\exp (-z/z_{0})
\end{equation}
With $z_{0} = 0.0417i-0.744$ (where the survey has sensitivity in the i band for magnitudes $21.5 < i < 23$).  In section~\ref{results} we will use LSST as an example next generation survey for constraints on parameters.  Choosing $i=22.25$ for a mid-range magnitude sample gives a fiducial $z_{0}=0.183$, which we use in our calculation.  The LSST science book assumes a galaxy bias evolution of b=1 + 0.84z.  Given a median redshift of around z=1, we will estimate the fiducial galaxy bias as b=1.84 in our simulations.  
Euclid is an ESA funded satellite mission due to launch in 2019.  It will observe 15,000 sq deg in optical and near-infrared using a 1.2m space telescope \cite{Laureijs:2011gra}.   
Figure~\ref{fig:dndz} shows the redshift distributions of these galaxy surveys along with the CMB lensing kernel.  Also shown is a Gaussian toy model with $\mu=1$ and $\sigma=0.5$.  This Gaussian is fairly representative of the galaxy distribution redshift ranges, and all redshift distributions show an overlap with the CMB lensing kernel, suggesting that they are useful for cross correlation studies.

In table~\ref{tab:scenarios} we show the combinations of CMB and galaxy surveys which we consider in our Fisher calculations, labelled from A-J.  To measure a cross correlation, the fractions of sky observed by CMB and galaxy surveys must overlap.  If the CMB experiment has a smaller $f_{\rm sky}$ than the galaxy survey, the smaller value must also be used for the galaxy information in that case.  We take a conservative approach, considering the same $f_{\rm sky}$ in the GG case as the $\phi G$ case, rather than using the full survey area available for GG.  These details can be read from column 3 of the table.  For the CMB information in the $TT,EE,TE$ correlations, we use the maximum $f_{\rm sky}$ possible for the deepest CMB survey (3rd or 4th generation).  We then include information from the current  ({\it Planck}) CMB survey, but only up to a combined $f_{\rm sky}=0.75$, since we do not want to double count patches of sky.  This counting assumes that the remaining 25\% of the sky are contaminated by foregrounds.  In scenarios E, F, G, H, I and J, we add 20\% priors to the galaxy parameters.  This is to break degeneracies between the galactic and non-galactic parameters in order to make the Fisher matrix invertible.  Adding e.g. 40\% priors rather than 20\% priors made negligible difference to the forecasts.

The GG and $\phi$G correlations have a sharper cut-off in multipole $l$, because non-linear clustering and scale dependence of the bias will make recovery of cosmological information from these scales challenging.  The approximate onset of non-linearities at $k\sim 0.2$ h/Mpc 
yields a projected $l_{\rm non-linear} \simeq 300$ in GG and $\phi$G correlations for a galaxy population peaking at $z\simeq1$.  The $\phi$G correlation should be less sensitive to the effect of non-linearities because it narrows in on higher redshifts. Taking this into consideration, we choose a multipole cut-off of $l_{\rm max}=500$.  We also explore the cases of $l_{\rm max}=1,000$ and $l_{\rm max}=5,000$ for this observable, assuming future improvements in our understanding of the nonlinear clustering regime.  Note that we have not modeled scale dependence in the bias due to non-linearities. Simulations show that non-linear bias should only affect scales $k\geq 0.7$ h/Mpc \cite{Zarija08}, which is outside the range of modes that we use as our default $l_{\rm max}$. However when we move beyond $l_{\rm max}=1,000$ we are entering this more complicated regime.

In some cases, we show forecasted constraints for a toy model Gaussian redshift distribution where we vary redshift-integrated galaxy density and $l_{\rm max}$.  Note that the {\it values} of number density, redshift distribution, bias, and $l_{\rm max}$ are not independent, as for example a change in the number density (by observing down to lower fluxes) would likely affect the mean galaxy bias. Since we are interested in forecasting parameter {\it errors} around a fiducial model, we can ignore such dependencies. 
Note also that we do not explicitly model a redshift dependence of the bias, b(z). This dependence is highly degenerate with the evolution of the number density N(z) when using observables integrated over all redshifts, as we do here. 
The degeneracy can be reduced by using redshift-binned galaxy auto and CMB lensing-galaxy cross power spectra.  We defer such an analysis to a future work. 

\begin{table}
\caption{CMB EXPERIMENTAL SCENARIOS}
\begin{tabular}{c | c c c}
& Current & 3rd gen. & 4th gen.   \\ \hline
temperature noise ($\mu$K-arcmin) & 30 & 2.5 & 2.5    \\ 
polarization noise ($\mu$K-arcmin)  & 60  & 3.5 & 3.5    \\ 
beam (arcmin) & 7  & 1 & 1    \\ 
$f_{\rm sky}$ & 0.75 & 0.1 & 0.5    \\ 
\end{tabular}
\label{CMBspecs}
\end{table}

\begin{table}
\caption{GALAXY EXPERIMENTAL SCENARIOS}
\begin{tabular}{c | c c c}
& $f_{\rm sky}$ & number density (gal/deg$^2$)   \\ \hline
WISE & 0.75 & 10,000   \\ 
LSST &0.5  & 198,000   \\  
EUCLID &0.4 & 108,000  \\ 
\end{tabular}
\label{GALspecs}
\end{table}

\begin{table*}
\caption{FISHER CALCULATION SCENARIOS}
\begin{center}
\begin{tabular}{c l l l l}
Scenario & CMB surveys  & $GG,\phi\phi,G\phi$ surveys  & Parameters varied (\% prior) \\ \hline

A &  {\it Planck}+4G & Gaussian, 4G &  $\Lambda$CDM$\nu$+bias+$\mu$+$\sigma$ \\ 

B & {\it Planck}+4G & LSST, 4G &  $\Lambda$CDM$\nu$+bias+$z_{0}$ \\ 

C & {\it Planck}+3G & Gaussian 1k gal/deg$^{2}$, 3G & $\Lambda$CDM$\nu$+bias+$\mu$ \\ 

D& {\it Planck}+4G & Gaussian 1k gal/deg$^{2}$, 4G &  $\Lambda$CDM$\nu$+bias+$\mu$ \\ 

E &  {\it Planck}+3G & LSST, 3G & $\Lambda$CDM$\nu$+bias(20\%) \\ 

F & {\it Planck}+4G & LSST, 4G  &  $\Lambda$CDM$\nu$+bias(20\%) \\ 

G & {\it Planck}+4G & Gaussian, 4G&  $\Lambda$CDM$\nu$+bias(20\%)+$\mu(20\%)$ \\ 

H & {\it Planck} & LSST, {\it Planck}  & $\Lambda$CDM$\nu$+bias(20\%)\\ 

I & {\it Planck} & LSST, {\it Planck}  &  $\Lambda$CDM+bias(20\%) \\ 

J & {\it Planck}+4G & LSST, 4G  &  $\Lambda$CDM+bias(20\%) \\ 

\end{tabular}
\end{center}
\label{tab:scenarios}
\begin{tablenotes}[para]
A description of the combinations of CMB and galaxy experimental scenarios we consider in our Fisher calculations. Column 2 lists surveys which are used in the $TT,EE,TE$ correlations.  3G denotes a third generation experiment with 10\% sky coverage, 4G a fourth generation experiment with 50\%. Column 3 details the surveys used for the $GG,\phi\phi,G\phi$ correlations, with the sky fraction given by that of the deeper experiment in column 2. LSST is estimated to have a number density of 198,000 gal/deg$^{2}$. For the Gaussian cases, either the number density is varied, or 1,000 gal/deg$^{2}$ is used.  In column 2, surveys are separated by addition signs.  This denotes that the Fisher information matrix was constructed for each of these surveys and then added together.  In column 3 the surveys are separated by commas. This denotes that information from both surveys was used in one Fisher matrix to construct auto and cross correlations between the CMB lensing and galaxy power spectra. The sky areas used are always those of the the highest sensitivity CMB experiment. Column 4 shows what parameters have been varied, where  $\Lambda$CDM$(\nu)$ was defined as $\Omega_{\rm b}h^{2}+\Omega_{\rm CDM}h^{2}+\Omega_{\lambda}+\tau+ n_{\rm s}+A_{\rm s} (+\Sigma m_\nu)$. Notice that priors of 20\% have been added to the galaxy parameters in scenarios E, F, G, H, I, and J.  These are added to break degeneracies between the galactic and non-galactic parameters in order to make the Fisher matrix invertible.   
\end{tablenotes}
\end{table*}

\section{Constraints on cosmological parameters}
\label{results}

In the following sections we refer to the {\it unlensed} primary CMB information using the notation CMB $ul$.  This term encompasses the $TT$ primary CMB temperature power spectrum, the $EE$ primary CMB E-mode information and the $TE$ cross correlation information.  Using unlensed spectra ensures that we do not over-count lensing information when the temperature and polarization power spectra are considered in conjunction with the reconstruction-based $\phi\phi$ and $\phi G$ spectra (see \cite{Schmittfull:2013uea,zahn} for more detailed treatments that quantify the shared information content between these estimators).  In some cases we consider the lensed theoretical power spectra for the primary CMB.  This is denoted by CMB $le$.

\subsection{Galaxy properties}
In this section we present constraints on parameters describing the large scale tracer structure probed by the cross correlation with CMB lensing, for both a toy model Gaussian case and for a specifically LSST parameterized distribution.  In all cases, we allow the full 7 non-galactic cosmology parameters to vary. In the case of the toy model describing the populations probed by the galaxy survey, we also vary 3 parameters describing the galaxy properties: the bias (b), and the mean ($\mu$) and standard deviation ($\sigma$) of the assumed Gaussian redshift distribution, shown by the black curve of figure~\ref{fig:dndz}.  

\begin{figure*}
\centerline{
\includegraphics[width=2.5in, angle=-90]{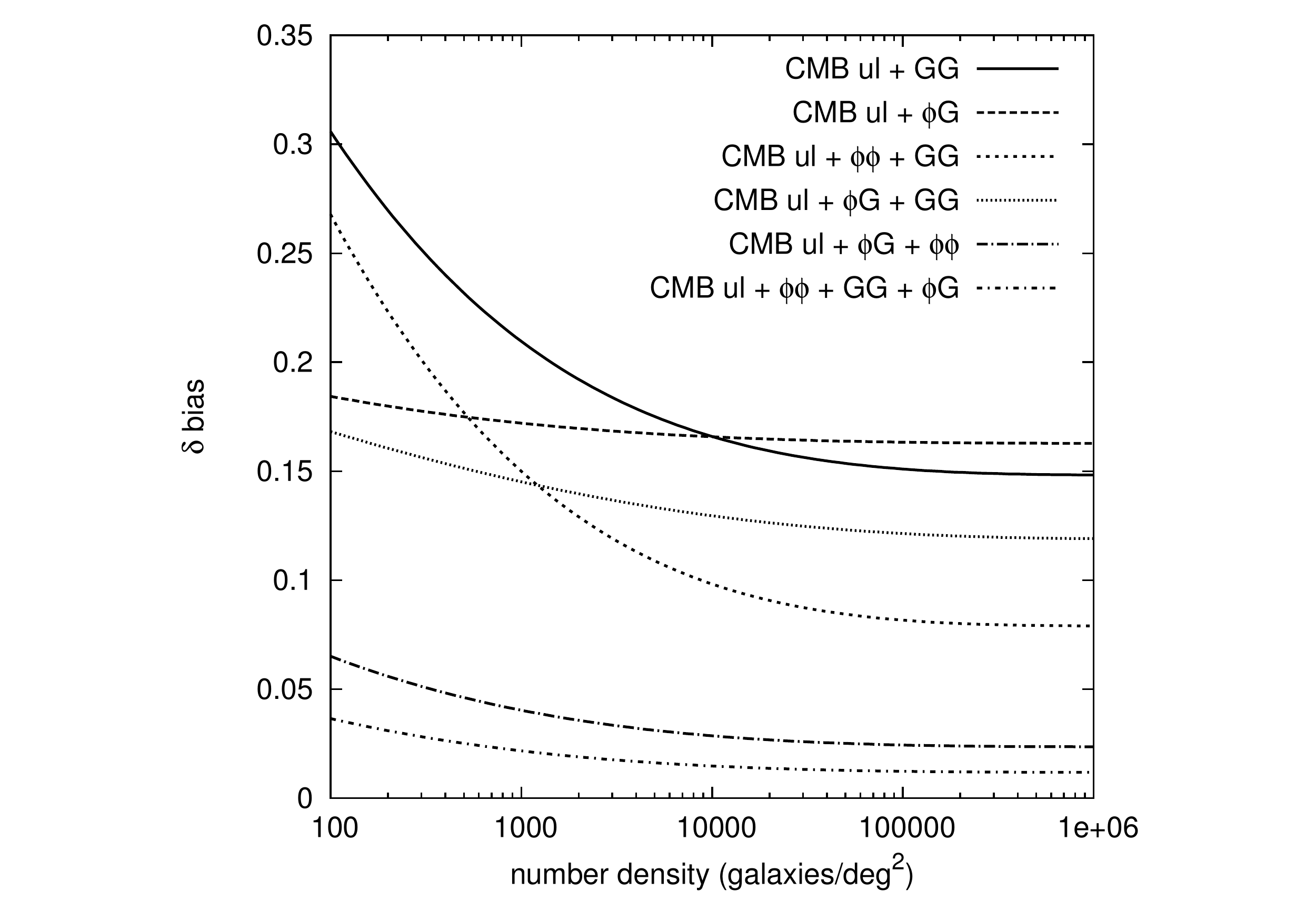}
\hspace{-1.1in}
\includegraphics[width=2.5in, angle=-90]{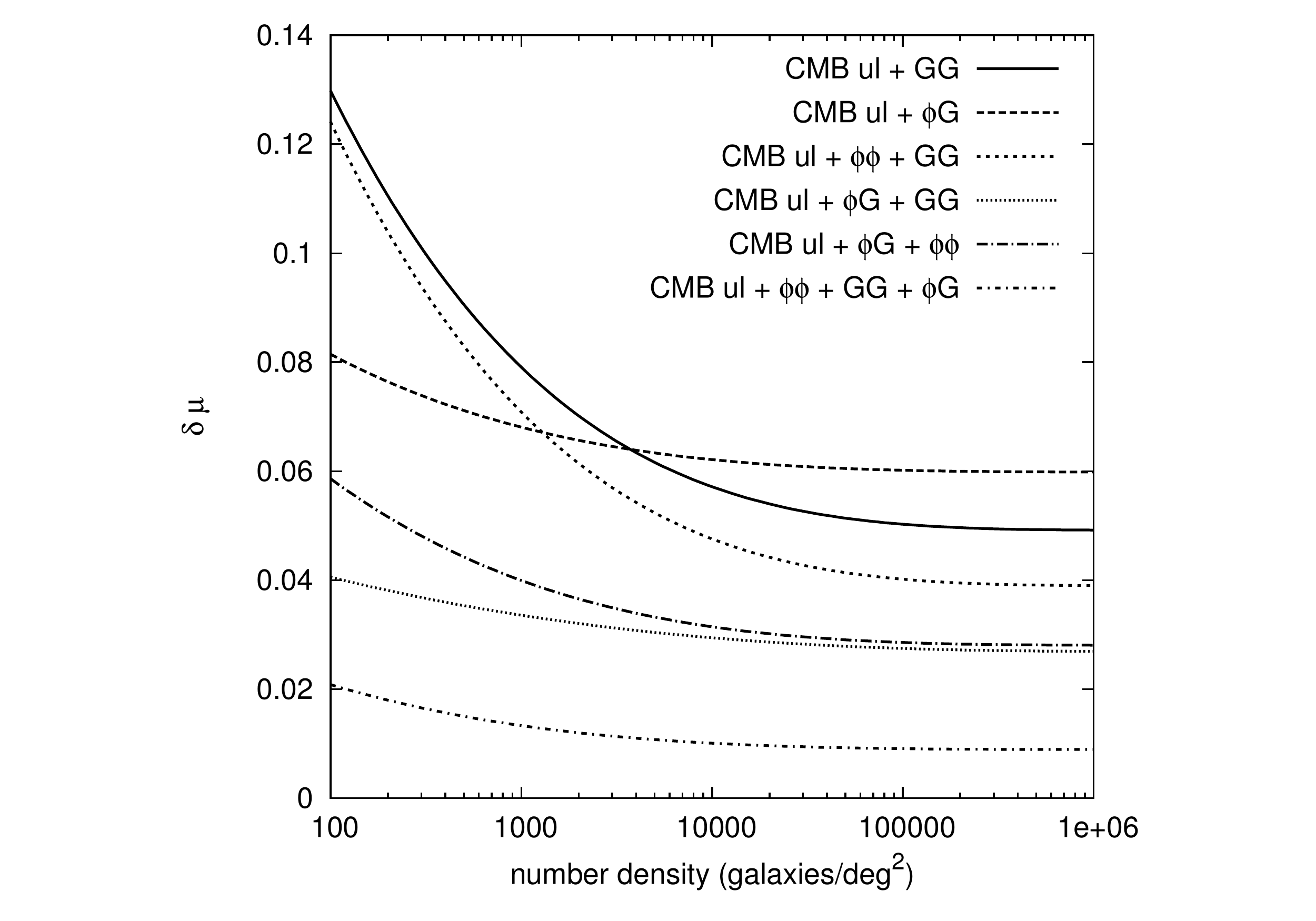}
\hspace{-1.1in}
\includegraphics[width=2.5in, angle=-90]{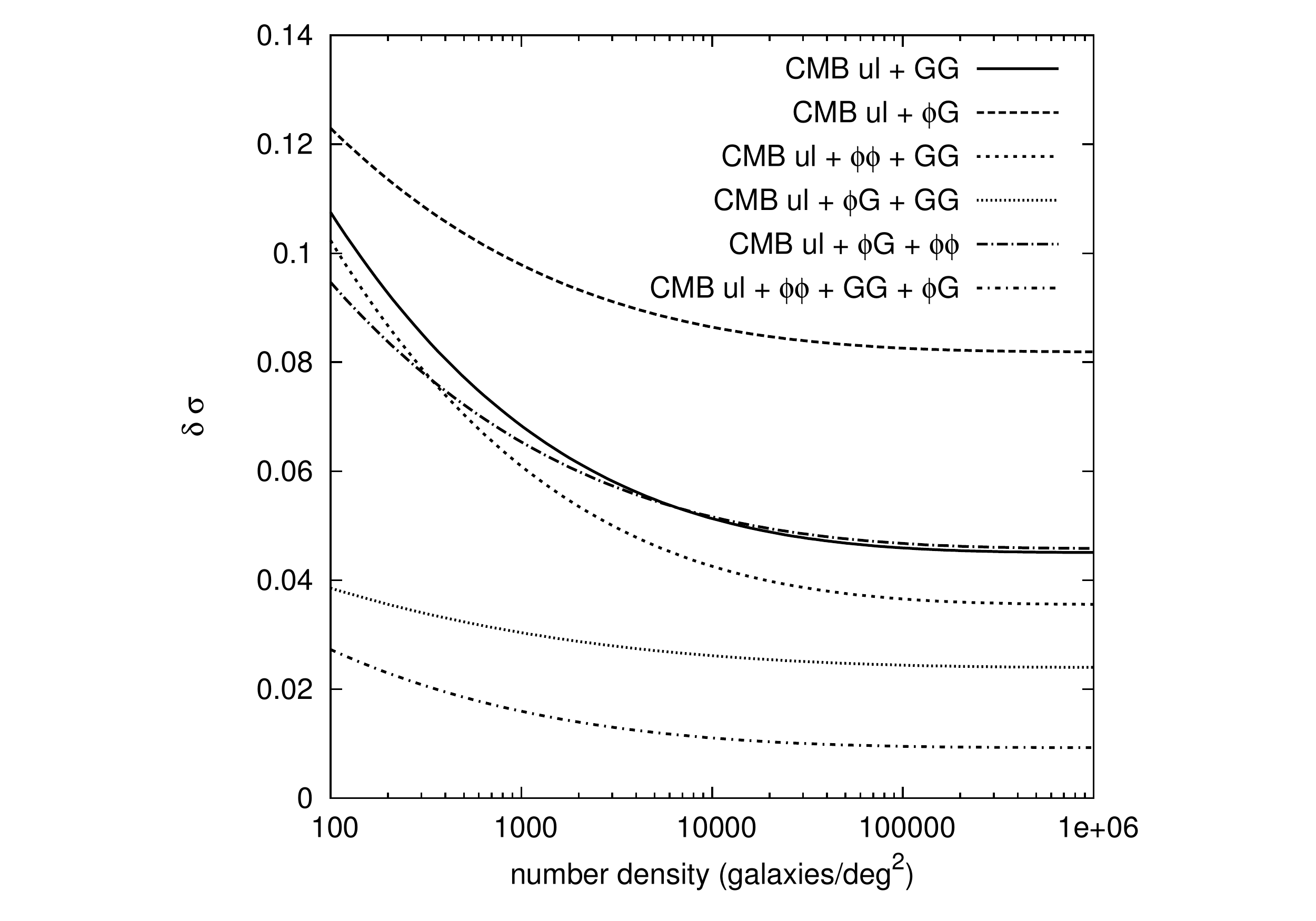}
}
\caption{
Constraints on the 3 galaxy parameters b, $\mu$ and $\sigma$ for a scenario-A experiment combination.   Left panel shows the galaxy bias parameter (b), centre panel shows the mean ($\mu$) of the Gaussian distribution, right panel shows the sigma ($\sigma$) of the Gaussian distribution.  The x axes show the number density of galaxies per square degree, and constraints are shown for various combinations of GG, $\phi\phi$ and $\phi$G with the primary CMB.}
\label{fig:gauss_dndz2}
\end{figure*}

Figure~\ref{fig:gauss_dndz2} shows constraints on b, $\mu$ and $\sigma$ for a scenario-A combination of surveys (a galaxy survey with a Gaussian redshift distribution and and \textit{Planck} + 4th. generation CMB experiments).  Parameter constraints are shown over a range of galaxy survey sensitivities (galaxy number density per square degree).  Different combinations of the lensing and galaxy auto correlations $\phi\phi$, GG, and the lensing galaxy cross correlation $\phi$G and the primary CMB are shown by lines on the plot.  For all three parameters, the addition of $\phi$G information improves constraints significantly over using CMB + GG alone (solid line to dotted line).  The constraints can be improved further by adding $\phi\phi$ information (bottom dot-dashed lines).  The addition of CMB lensing information is crucial for constraining galaxy bias, because it helps break degeneracies between bias and other cosmological parameters affecting the amplitude and shape of the galaxy power spectra.  

\begin{table}[b]
\caption{}
\begin{center}
\begin{tabular}{l | c |c }
scenario-B& $\delta$z$_{0}$ & $\delta$b    \\ \hline
CMB ul +GG  &  0.010   & 0.065  \\ 
CMB ul + $\phi$G  &  0.015   & 0.093   \\ 
CMB ul + $\phi\phi$ + $\phi$G  &  0.010  &   0.017   \\ 
CMB ul + $\phi$G + GG &  0.002     &  0.040   \\
CMB ul + $\phi\phi$ + GG &  0.007   &  0.060    \\  
CMB ul + $\phi\phi$ + GG + $\phi$G &  0.002   &   0.016   \\ 
\end{tabular}
\label{tab:dndzLSST2}
\begin{tablenotes}[para]
Constraints on the galaxy parameters b and $z_{0}$ for a scenario-B experiment combination.  (LSST galaxy survey with \textit{Planck} + 4th generation CMB)  Values are shown for different combinations of GG, $\phi\phi$ and $\phi$G with the primary CMB.  
\end{tablenotes}
\end{center}

\end{table}

Table~\ref{tab:dndzLSST2} shows constraints on galaxy parameters b and $z_{0}$ for an LSST type survey (described in section~\ref{data} and shown in equation~\ref{LSST}), for the combination with a 4th generation CMB experiment and {\it Planck}.  $z_{0}$ is the single parameter describing the shape of the redshift distribution.  When the cross correlation is added to the autocorrelations (CMB ul + $\phi\phi$ + GG case to CMB ul + $\phi\phi$ + GG + $\phi G$ case), the constraints on both the bias and shape parameter improve by $\sim$70\%.  
Interestingly, the constraints on both parameters from CMB ul + GG is only 30\% improved from the constraint of CMB ul + $\phi$G.  This shows that the CMB lensing-galaxy correlation contains powerful information on galaxy dynamics, in particular the bias, even without using the galaxy clustering auto power spectrum. The significant improvement in the constraint on the shape parameter $z_0$ in the combination of all observables (last line) compared to CMB ul + GG or CMB ul + $\phi$G + $\phi\phi$ shows on the other hand that the galaxy clustering auto correlation becomes much more useful with the addition of CMB lensing information. 

\begin{figure*}
\begin{center}
\subfloat[Scenario-C]{\includegraphics[width=0.44\textwidth, height=0.25\textheight]{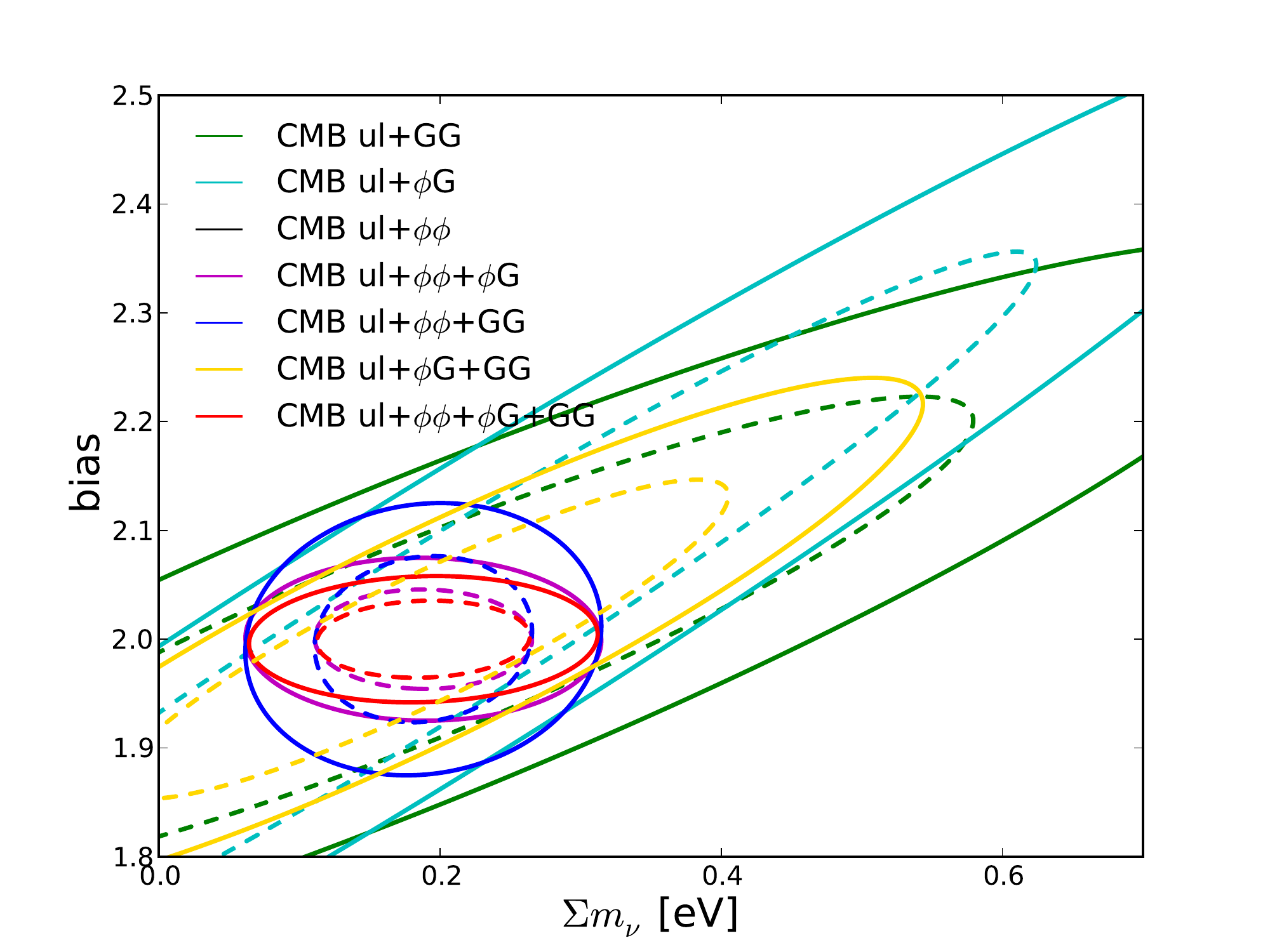}} 
\label{subfig:nu_vs_bias_3g}
\subfloat[Scenario-D ]{\includegraphics[width=0.44\textwidth,height=0.25\textheight]{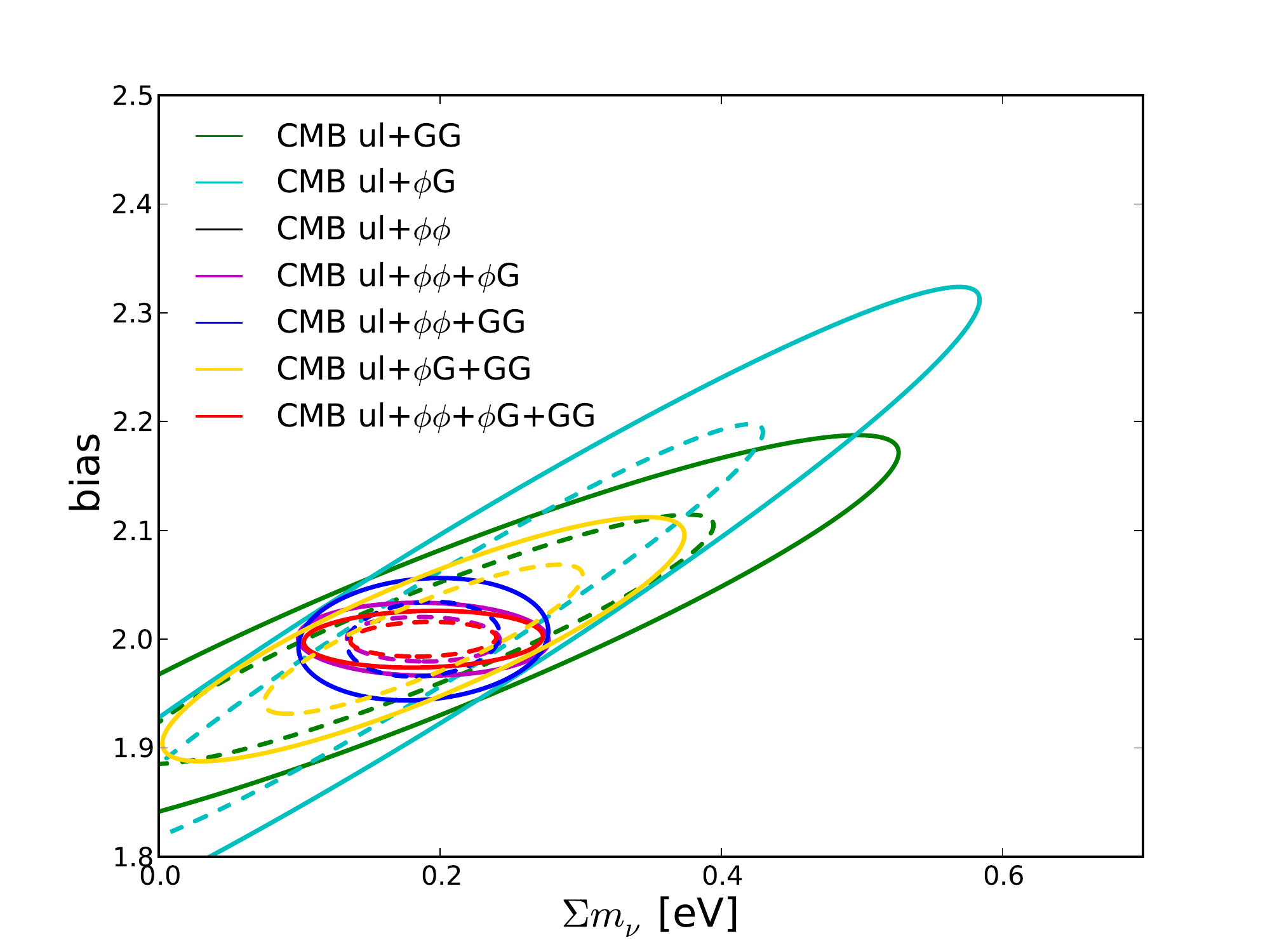}}
\vspace{-0.4cm}
\noindent 
\subfloat[Scenario-C]{\includegraphics[width=0.44\textwidth, height=0.25\textheight]{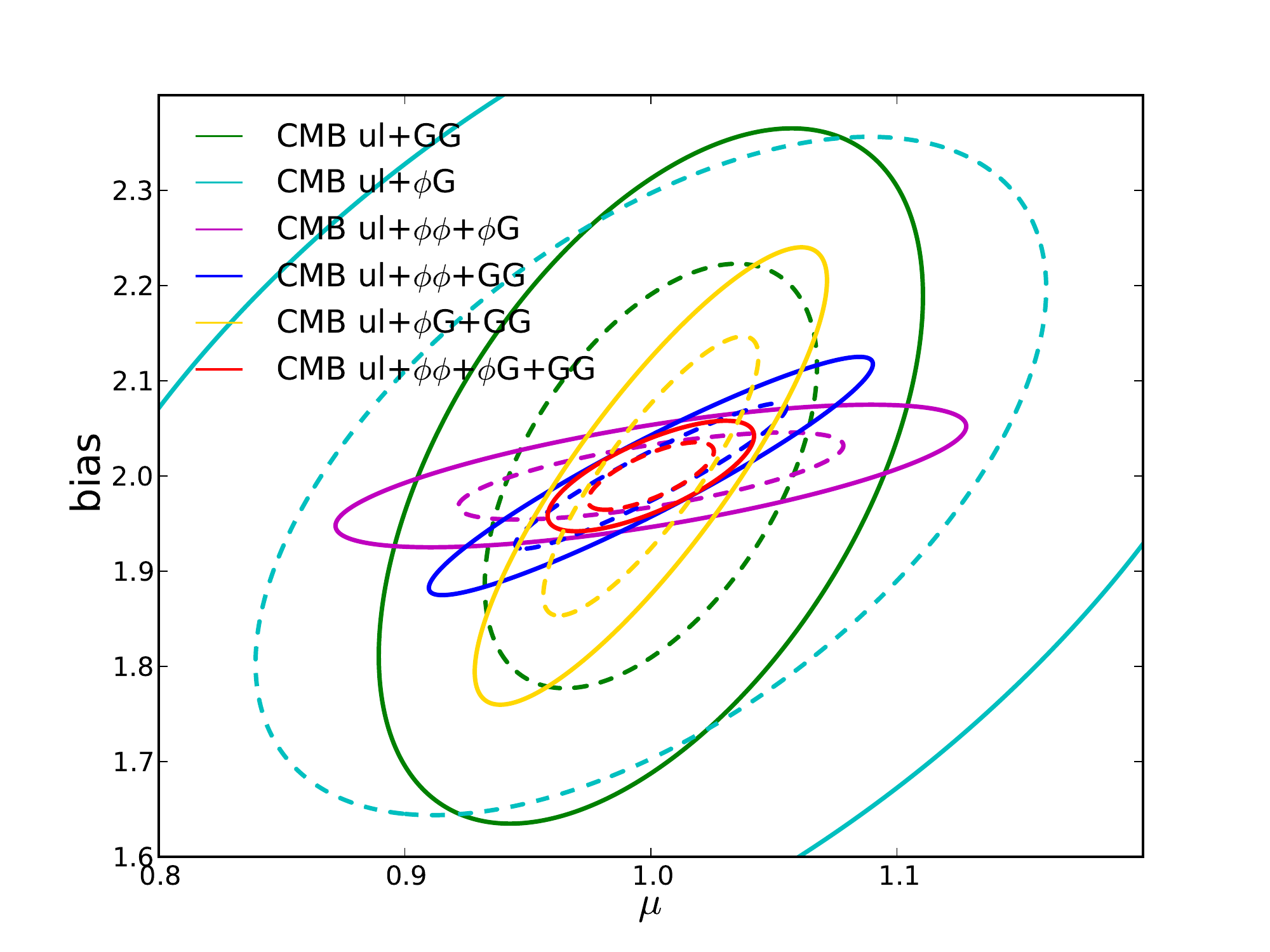}}
\subfloat[Scenario-D]{\includegraphics[width=0.44\textwidth,height=0.25\textheight]{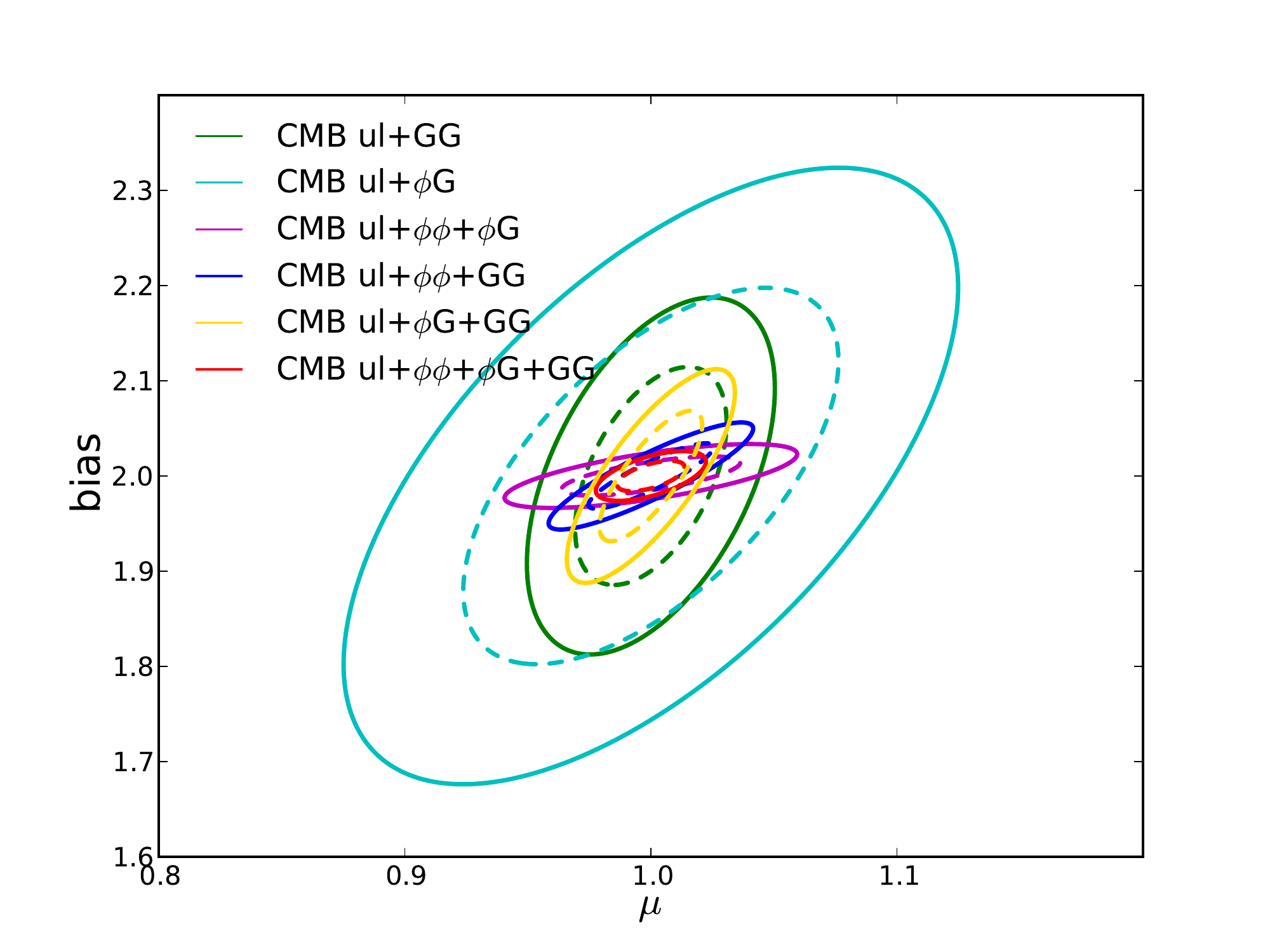}}
\caption[caption]{
1$\sigma$ (dashed line) and 2$\sigma$ (solid line) ellipses for various combinations of GG, $\phi\phi$ and $\phi$G with the primary CMB.  Left and right panels show the comparison between a scenario-C (3rd generation CMB) and scenario-D (4th generation CMB) experiment combination.  Galaxy survey is Gaussian with number density 1,000 gal/deg$^{2}$.} 
\label{fig:astro_ellipse}
\end{center}
\end{figure*}

Figure~\ref{fig:astro_ellipse} shows the 1$\sigma$ and 2$\sigma$ ellipses for various parameters for scenario-C and scenario-D experiment combinations.  (The galaxy survey is Gaussian with a number density of 1,000 gal/deg$^{2}$.  The difference between scenario-C and scenario-D is that the former includes a 3rd generation CMB survey, whilst the latter includes a 4th generation CMB survey.)  The top two panels show a comparison of the $\sum m_{\nu}$ and b parameters for both scenario-C (left) and scenario-D (right).  The two parameters are highly degenerate when only including the galaxy auto power or galaxy-lensing cross correlation. We see again that CMB ul + $\phi$G combination does almost as well as the CMB ul + GG combination.  The improvement in the CMB ul + $\phi$G cross combination is $\sim$50\% for both parameters when upgrading to scenario-D from scenario-C.  The lower two panels show a comparison of the redshift distribution mean ($\mu$) and bias parameter for scenario-C (left) and scenario-D (right).  The improvement in the $\mu$ parameter is $\sim$40\% when upgrading to scenario-D from scenario-C.    In all cases we see that adding the $\phi\phi$ lensing auto correlation to the CMB ul +GG (green line to blue line) or CMB ul + $\phi$G (cyan line to purple line) improves the constraints significantly, because the lensing auto power constrains neutrino mass more directly.  There is also a good improvement in the constraints when $\phi G$ is added to the CMB ul + GG (green line to yellow line), although the improvement is more dramatic with the addition of the lensing autocorrelation.  In both choices of parameter pairs, when comparing scenario-C and scenario-D, the CMB ul +GG constraints improve about as much as the CMB ul + $\phi$G constraints.  

\subsection{Massive neutrinos and reionization optical depth}

\begin{figure}
\begin{center}
\hspace*{-0.2cm}
\includegraphics[width=8.8cm]{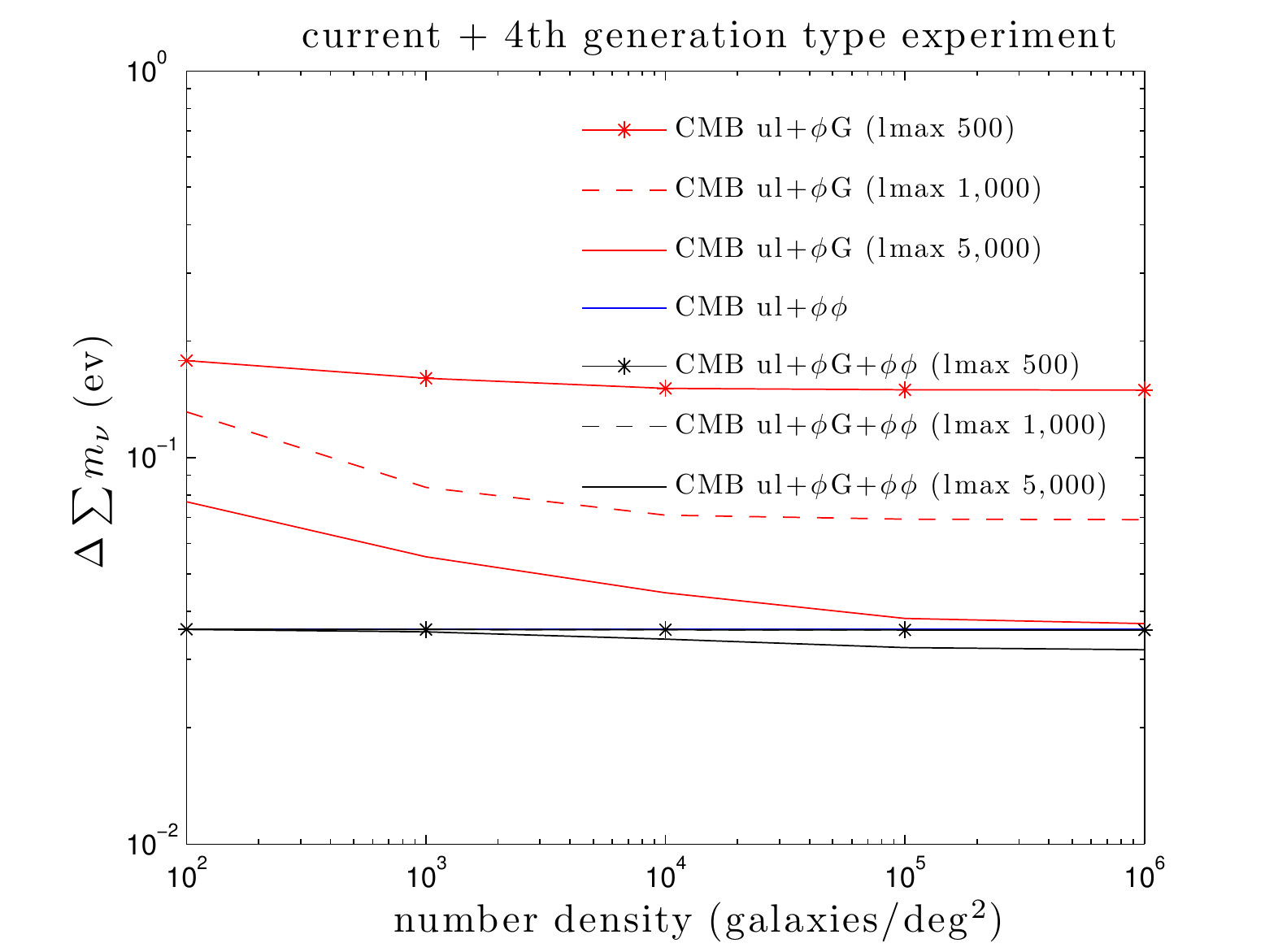}
\caption{$\Delta\sum m_{\nu}$ for scenario-G.  Constraints in the y axis are shown in electron volts, the x axis shows number density of galaxies observed per square degree.  Forecasts are shown for different cutoff values of multipole $l_{\rm max}$=500, 1,000, 5,000.}
\label{fig:lmax_compare}
\end{center}
\end{figure}

In this section we present forecasts on the non-galactic cosmological parameters: the neutrino mass sum $\sum m_{\nu}$, and the optical depth $\tau$ due to free electrons since the epoch of reionization.  

Figure~\ref{fig:lmax_compare} shows the $\sum m_{\nu}$ constraint for scenario-G for different cutoff values, $l_{\rm max}=500$, $l_{\rm max}=1,000$, $l_{\rm max}=5,000$.  For each value of $l_{\rm max}$ the primary CMB is shown in combination with the $\phi G$ cross correlation (red lines), and the cross correlation plus lensing auto-correlation $\phi G+\phi\phi$ (black lines).  These lines can be compared to that of the primary CMB plus only the CMB lensing auto-correlation $\phi\phi$ (blue line).  (Remember that the primary CMB is unlensed and includes the TT, EE and TE correlations).  In this scenario, the $\sigma$ of the redshift distribution was fixed, but the b and $\mu$ parameters were allowed to vary with a 20\% prior in order to allow matrix inversions.  Using a prior of e.g., 40\% made negligible difference to the constraints, as did fixing the $\mu$ parameter rather than varying it.  We see that adding the $\phi$G information to the CMB ul + $\phi\phi$ combination (blue line to black lines) leads to negligible improvement unless one allows $l_{\rm max}=5,000$.  For a conservative value of $l_{\rm max}=500$, the combination of the unlensed CMB and $\phi$G does not give useful constraints.  If we consider a less conservative $l_{\rm max}=1,000$,  the CMB ul + $\phi$G combination becomes more valuable, and if we consider an optimistic $l_{\rm max}=5,000$ then the constraint becomes comparable to that of the CMB ul + $\phi\phi$ line.  With optimistic $l_{\rm max}$ limits, the $\phi G$ cross correlation could provide a good independent check on the $\sum m_{\nu}$ parameter.  It is worth mentioning that while the CMB ul + $\phi$G combination places better constraints than the CMB ul does alone, what we actually measure is the lensed primary CMB, and this constrains the $\sum m_{\nu}$ much better than the cross correlation and almost as well as the CMB lensing autocorrelation.  (When we consider the $\phi\phi$ correlation in combination with the primary CMB, we always use the unlensed CMB to avoid over-counting the lensing effect).

\begin{table}[b]
\caption{}
\begin{adjustwidth}{-0.5cm}{}
\begin{tabular}{c| c |c }
& $\Delta \sum m_{\nu}$ (scenario-E) & $\Delta \sum m_{\nu}$ (scenario-F)  \\ \hline
CMB ul only & 0.328  & 0.201  \\ 
CMB  le &0.055  &  0.040  \\ 
CMB ul + $\phi\phi$  & 0.051 &   0.036     \\ 
CMB ul + GG  & 0.122 & 0.059       \\ 
CMB ul + $\phi$G  & 0.219  & 0.119      \\ 
CMB ul + $\phi\phi$ + $\phi$G  & 0.051 &   0.035   \\ 
CMB ul + GG + $\phi$G  & 0.114 & 0.058     \\ 
CMB ul + $\phi\phi$ + GG   & 0.050 & 0.034     \\ 
CMB ul + $\phi\phi$ + GG +$\phi$G  &0.044  & 0.027     \\ 
\end{tabular}
\end{adjustwidth}
\label{LSSTnu}
\begin{tablenotes}[para]
Constraints in electron volts on the $\sum m_{\nu}$ parameter for an LSST type galaxy survey (equation~\ref{LSST}) for scenario-E (3rd generation CMB) and scenario-F (4th generation CMB).  Constraints are shown for combinations of GG, $\phi\phi$ and $\phi$G with the primary CMB. 
\end{tablenotes}
\end{table}

Tables~\ref{LSSTnu} and~\ref{LSSTtau} show constraints on the $\sum m_{\nu}$ and $\tau$ parameters for an  LSST type galaxy survey (described in equation~\ref{LSST}).  In this case the bias parameter was set to a fiducial value of b=1.84.  

Table~\ref{LSSTnu} shows constraints on $\sum m_{\nu}$ for scenario-E and scenario-F.  The only difference between scenario-E and scenario-F is that the former includes a 3rd generation CMB experiment, while the latter includes a 4th generation CMB experiment.  In the same trend as the Gaussian redshift distribution galaxy survey, using the lensed CMB vs the unlensed CMB gives dramatic improvements, almost an order of magnitude in this case.  The lensed CMB alone is much better than the CMB ul + $\phi$G, and the addition of $\phi$G to CMB ul + $\phi\phi$ gives negligible improvement.  However, the improvement in the neutrino constraint from CMB ul + GG + $\phi\phi$  to CMB ul + GG + $\phi\phi$ + $\phi G$ is 10\% for a 3rd generation CMB experiment and 20\% for a 4th generation.  This shows that when the redshift distribution is well known, the addition of $\phi G$ helps break degeneracies between cosmological and galaxy parameters in GG.  

\begin{table}[b]
\caption{}
\begin{center}
\begin{tabular}{l | c |c }
$\Lambda$CDM$\nu$& $\Delta \tau$ (scenario-H) & $\Delta \tau$(scenario-F)   \\ \hline
TT len  & 0.084  & 0.036 \\
CMB le  & 0.004  & 0.003 \\
TT+$\phi\phi$ & 0.041  & 0.015 \\
TT+GG & 0.064  & 0.056 \\
TT+$\phi$G   &  0.068& 0.055 \\
TT+$\phi\phi$+$\phi$G  &  0.039  & 0.014  \\ 
TT+GG+$\phi$G  & 0.037   & 0.025 \\
TT+GG+$\phi\phi$ & 0.028   & 0.015 \\
TT+GG+$\phi\phi$ +$\phi$G&  0.028  & 0.014 \\ \hline

 $\Lambda$CDM & $\Delta \tau$ (scenario-I) & $\Delta \tau$(scenario-J)   \\ \hline
TT len  & 0.027 & 0.013 \\ 
CMB le  & 0.004  & 0.003 \\ 
TT+$\phi\phi$  &0.020  & 0.010 \\ 
TT+GG & 0.019 & 0.018   \\
TT+$\phi$G  & 0.047 & 0.033  \\ 
TT+$\phi\phi$+$\phi$G  & 0.020 & 0.010 \\
TT+GG+$\phi$G  & 0.019   & 0.016 \\
TT+GG+$\phi\phi$  & 0.017  & 0.010  \\
TT+GG+$\phi\phi$ +$\phi$G & 0.016   & 0.007 \\ \hline

\end{tabular}
\end{center}
\label{LSSTtau}
\begin{tablenotes}[para]
Constraints on the $\tau$ parameter for an LSST type galaxy survey (equation~\ref{LSST}).  Constraints are shown for a number of scenarios for combinations of TT, CMB, $\phi\phi$, GG and $\phi$G.  Scenarios H and I are \textit{Planck} like, scenarios F and J are 4th generation CMB like.
\end{tablenotes}
\end{table}

In table~\ref{LSSTtau}, we show the constraint on $\tau$ for various scenarios.  Scenario-H and scenario-F use the full 7 parameter non-galactic cosmology where scenario-H includes {\it Planck} and scenario-F includes a 4th generation CMB survey.  Also shown are scenario-I and scenario-J.  These two scenarios include only the $\Lambda$CDM 6 non-galactic cosmology parameters, excluding massive neutrinos.  Scenario-I includes a current type CMB survey and scenario-J includes a 4th generation CMB survey.  When the neutrino mass sum is fixed, the constraint on $\tau$ improves significantly for many cases.  This highlights the importance of including the neutrino mass sum as an unknown parameter in the analysis. 

While $\phi$G reduces degeneracies between cosmological parameters inherent in TT+GG, the improvement in the tau constraint going from TT + GG + $\phi\phi$  to TT + GG + $\phi\phi$ + $\phi G$ is small.  Note that while large scale polarization information from {\it Planck} or future experiments will yield the tightest constraints on $\tau$, the addition of $\phi\phi$ information alone to the temperature power spectrum (TT) improves the $\tau$ constraints significantly.  
The constraint from a 4th generation CMB experiment {\it without} large scale polarization information will be comparable to that from current large scale polarization measurements from WMAP \cite{Hinshaw:2012aka} and will offer a good independent check on the constraint from Planck polarization, to be published next year, which could be dominated by uncertainty in the polarization foregrounds. 
  
\begin{figure}
\begin{center}
\hspace*{-0.2cm}
\includegraphics[width=8.8cm]{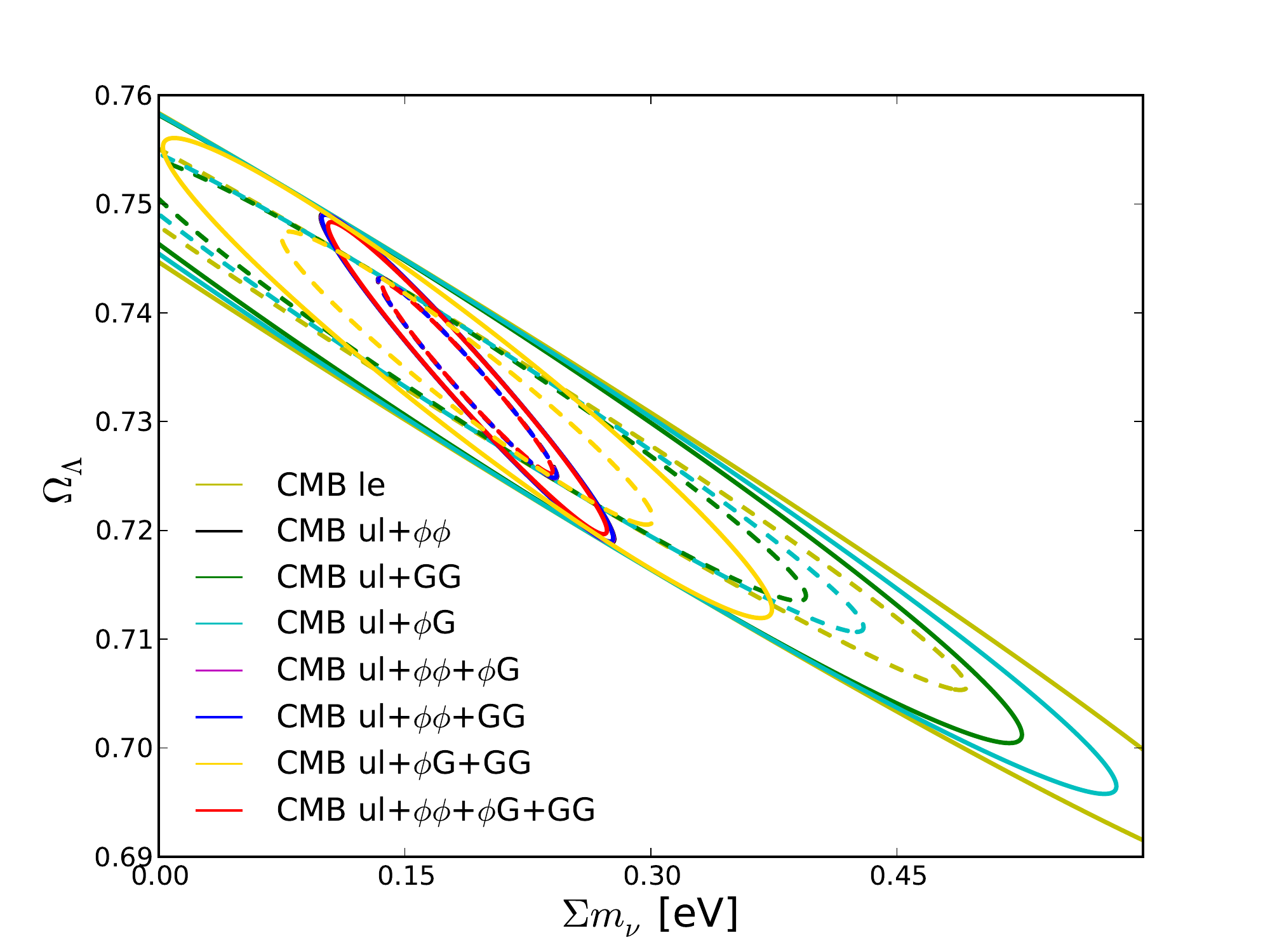}
\caption{$\sum m_{\nu}$ vs $\Omega_{\Lambda}$ 1$\sigma$ (dashed line) and 2$\sigma$ (solid line) ellipses for scenario-D for various combination of the CMB ul, CMB lensing, galaxy and cross correlations.  Galaxy survey is Gaussian with number density 1,000 gal/deg$^{2}$, and CMB experiment is 4th generation.}
\label{fig:ellipse_nu_vs_lam}
\end{center}
\end{figure}

Figure~\ref{fig:ellipse_nu_vs_lam} shows the 1$\sigma$ and 2$\sigma$, $\sum m_{\nu}$ vs $\Omega_{\Lambda}$ ellipses for scenario-D.  (Galaxy survey with a Gaussian galaxy distribution and a number density of 1,000 gal/deg$^{2}$, \textit{Planck} + 4th generation CMB experiments).  As expected, adding the CMB lensing auto correlation to the primary CMB improves the constraint dramatically, due to the breaking of the geometric degeneracy (\cite{Stompor:1998zj}, \cite{Sherwin:2011gv}).  However, after the addition of the $\phi\phi$ correlation, additional information from both GG and $\phi$G make negligible improvement to the constraint on dark energy.  In the absence of the $\phi\phi$ correlation, the addition of  GG or $\phi$G also make negligible improvement on the constraint from the primary CMB alone.

\section{Conclusions}
\label{conclusion}

Our analysis is intended to begin to answer the open question of how much information about cosmology is contained in the CMB lensing - galaxy cross correlation.  We have used a simple Fisher model where we included all correlations between temperature, polarization, galaxy over-density, and CMB lensing maps. We constrain both galactic and non-galactic cosmological parameters simultaneously, to explore degeneracies in these spectra between the underlying cosmology and galaxy dynamics.  

In the case of the galaxy parameters, the $\phi$G cross correlation is very useful.  In all cases the addition of the $\phi G$ cross correlation to the $\phi\phi$ and GG auto correlations improves the constraints significantly, and CMB ul + $\phi G$ is comparable to CMB ul + GG.  We also find that the addition of CMB lensing information or any improvements in the primary CMB significantly improve the constraints on galaxy parameters.  This is because the CMB lensing and primary CMB information give tighter constraints on the non-galactic cosmological parameters, allowing degeneracies with galaxy parameters to be broken.   

For non-galactic cosmology, we find that in the case of the $\sum m_{\nu}$, once the $\phi\phi$ and GG auto correlations are considered, the $\phi$G cross correlation adds little improvement.  We have imposed a conservative $l_{\rm max}=500$ for the $\phi$G cross correlation however, and see that when this is relaxed, the cross correlation does offer useful information, up to the same level as the auto correlations for ambitious values of $l_{\rm max}$.  Although for now it may be unrealistic to extend the cross correlation information to such small scales, with some improvement in the non-linear modeling one could hope that the cross correlation provides useful independent checks of constraints derived from the GG auto-correlation.  In table~\ref{LSSTtau} we also quantify the near future constraints available on the $\tau$ parameter when the CMB lensing information is used.  We note that the TT + $\phi\phi$ correlation will constrain $\tau$ to $<$20\%, which will provide a good independent check on the epoch of reionization,  given galactic foreground uncertainties in the polarization constraints.

\begin{acknowledgements}
The authors thank (in alphabetical order) Carlos Cunha,  Sudeep Das, Gill Holder, Antony Lewis, Adam Lidz, Blake Sherwin, Alberto Vallinotto, Kimmy Wu, and Amanda Yoho for useful discussions and comments on a draft.  RP acknowledges support from the Science and Technology Facilities Council via a research studentship and thanks Prof. Chao-Lin Kuo's group at SLAC/Stanford where they were hosted at the time of this work. OZ acknowledges support by an Inaugural Fellowship from the Berkeley Center for Cosmological Physics as well as by the National Science Foundation through grants ANT-0638937 and ANT-0130612.
\end{acknowledgements}

\providecommand{\aj}{Astron. J. }\providecommand{\apj}{Astrophys. J.
  }\providecommand{\apjl}{Astrophys. J.
  }\providecommand{\mnras}{MNRAS}\providecommand{\aap}{Astron.
  Astrophys.}\providecommand{\aj}{Astron. J. }\providecommand{\apj}{Astrophys.
  J. }\providecommand{\apjl}{Astrophys. J.
  }\providecommand{\mnras}{MNRAS}\providecommand{\aap}{Astron. Astrophys.}


\end{document}